\documentclass[aps,prl,twocolumn,showpacs,floatfix]{revtex4}
\usepackage{graphicx,color}
\usepackage{amsmath,amssymb}

\newcommand{\figmodel}
{\begin{figure}[htbp]
        \centering
        \includegraphics[height=1.in]{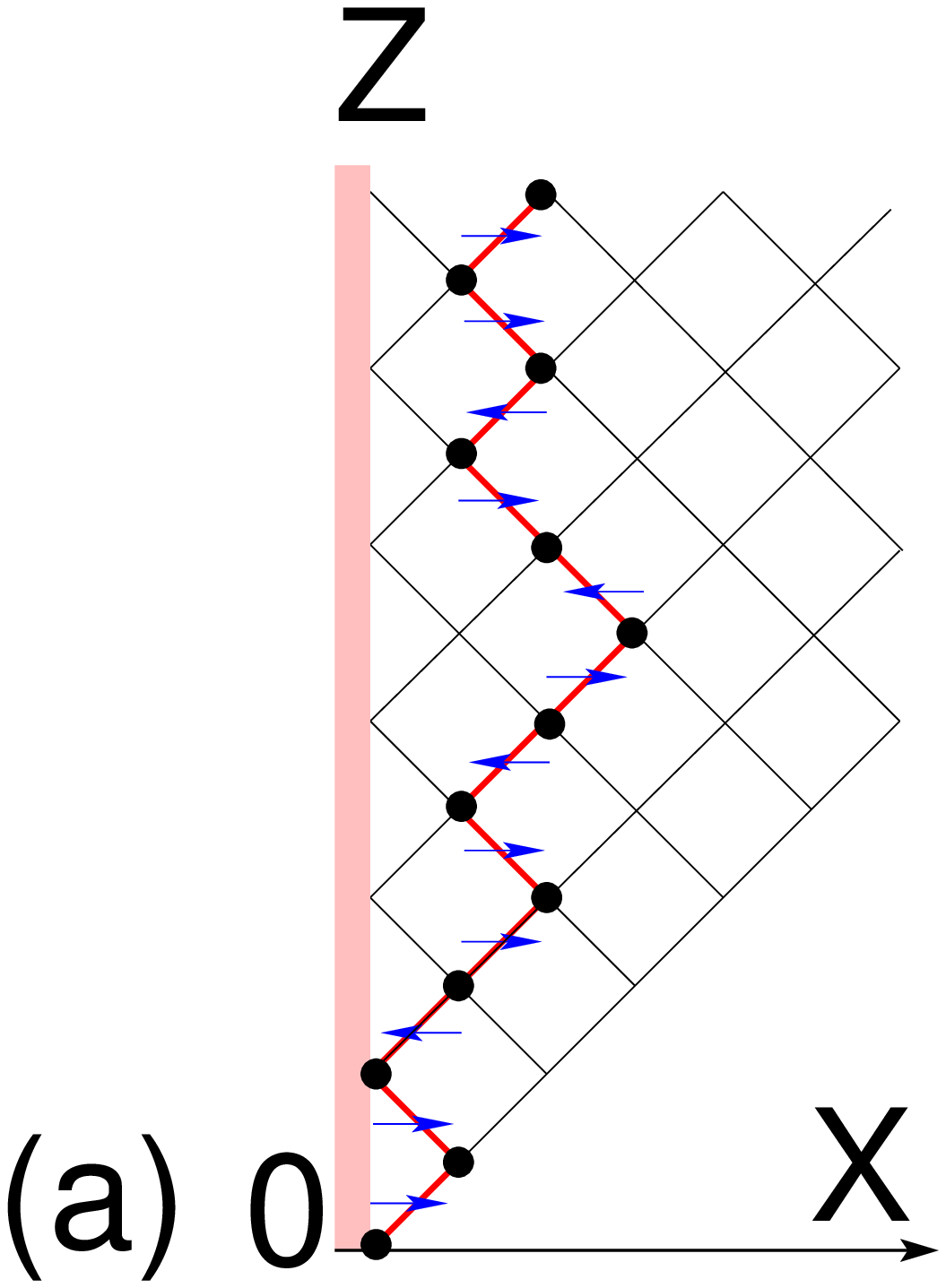}
		\hspace{0.2cm}
        \includegraphics[clip,height=1in]{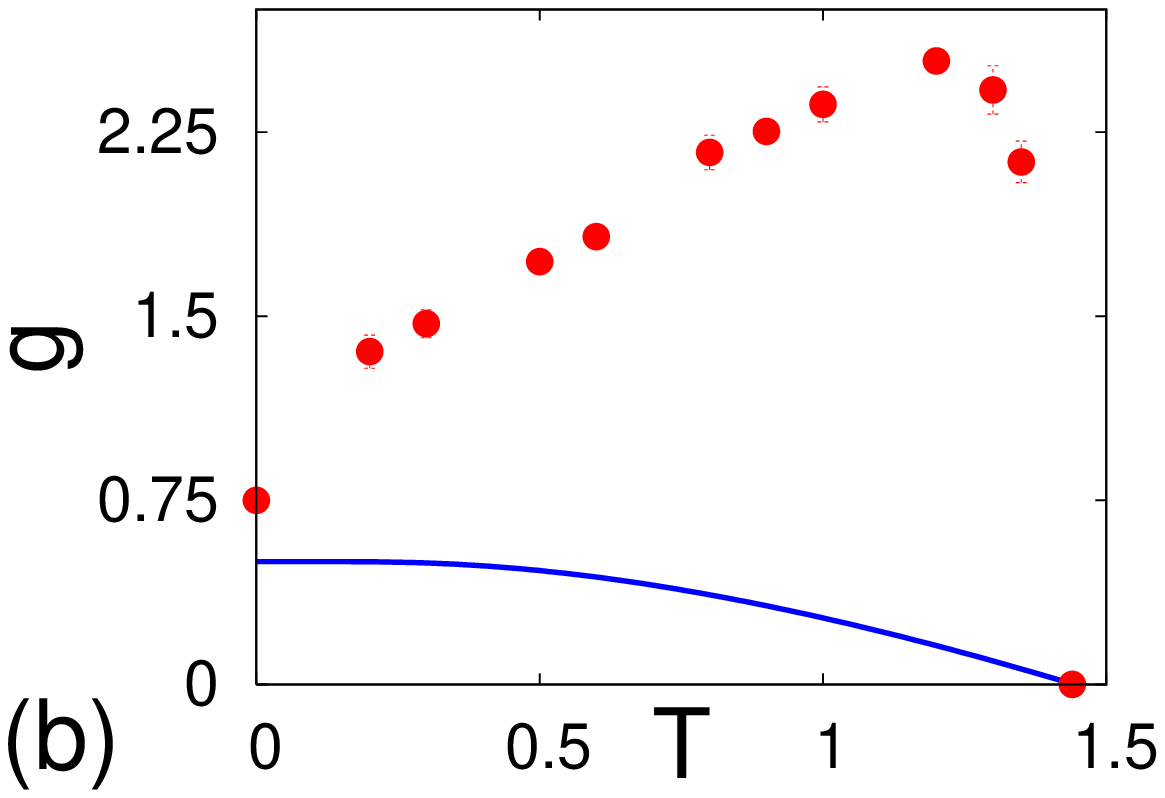}
        \caption{\label{fig:model} 
          (a) Schematic diagram of a directed polymer in $D = 1 + 1$
          dimensions with an attractive wall at $x = 0$. The direction
		  of the random force $g {\bf \hat{\zeta}}(z)$ is shown by the
		  arrows on each bond. (b) $g$ (in units of $\epsilon$) vs $T$
		  (in units of $k_B/\epsilon$) phase diagram. The points
		  are for the randomly forced polymer and solid curve is for the
		  pure case. The same convention is used in all the plots. 		  
		  }
\end{figure}
}

\newcommand{\figgs}
{\begin{figure}[htbp]
        \centering
        \includegraphics[width=1.4in]{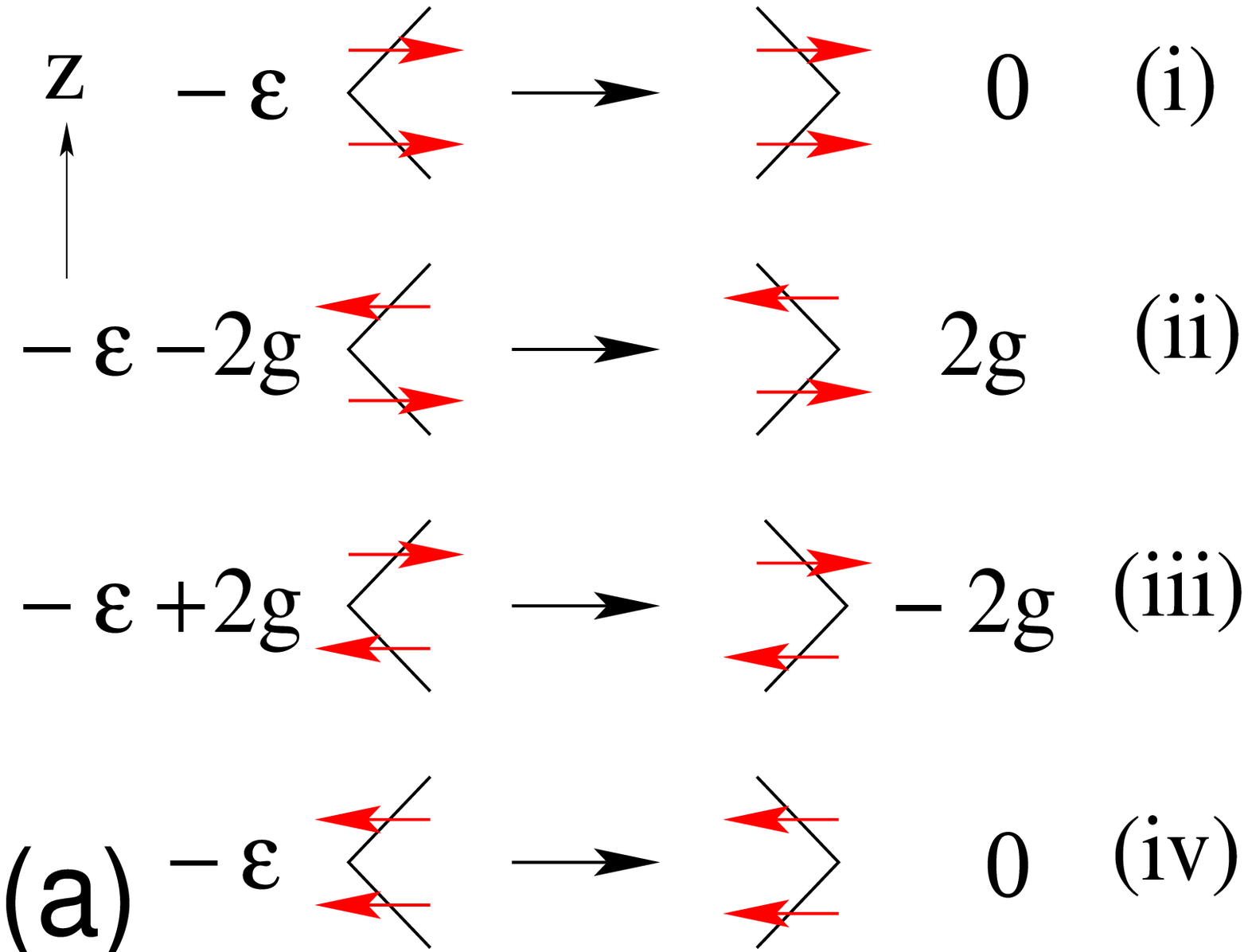}
		\hspace{0.2cm}
        \includegraphics[width=1.65in]{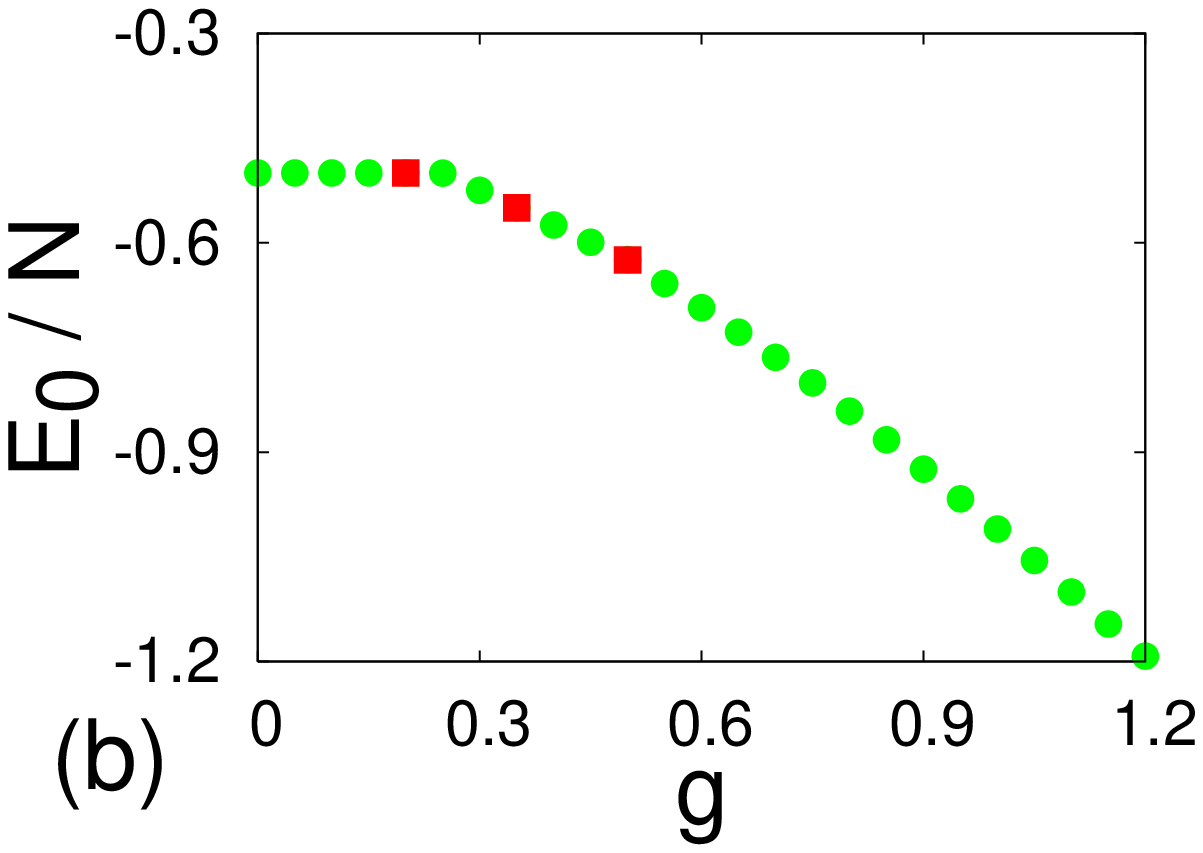}
        \caption{\label{fig:gs} 
          (a) The energy cost in flipping a monomer on the wall.
          (b) Ground state energy per monomer, $E_0/N$, as a function
          of force obtained for $N=256$ using exact transfer matrix at
          $T=0.001$ (circles). Estimates from single flip Monte Carlo
          at $T=0$ for $N=1024$ (square) ($\epsilon=1$).}
\end{figure}
}

\newcommand{\figgx}
{\begin{figure}[htbp]
        \centering
        \includegraphics[width=1.6in]{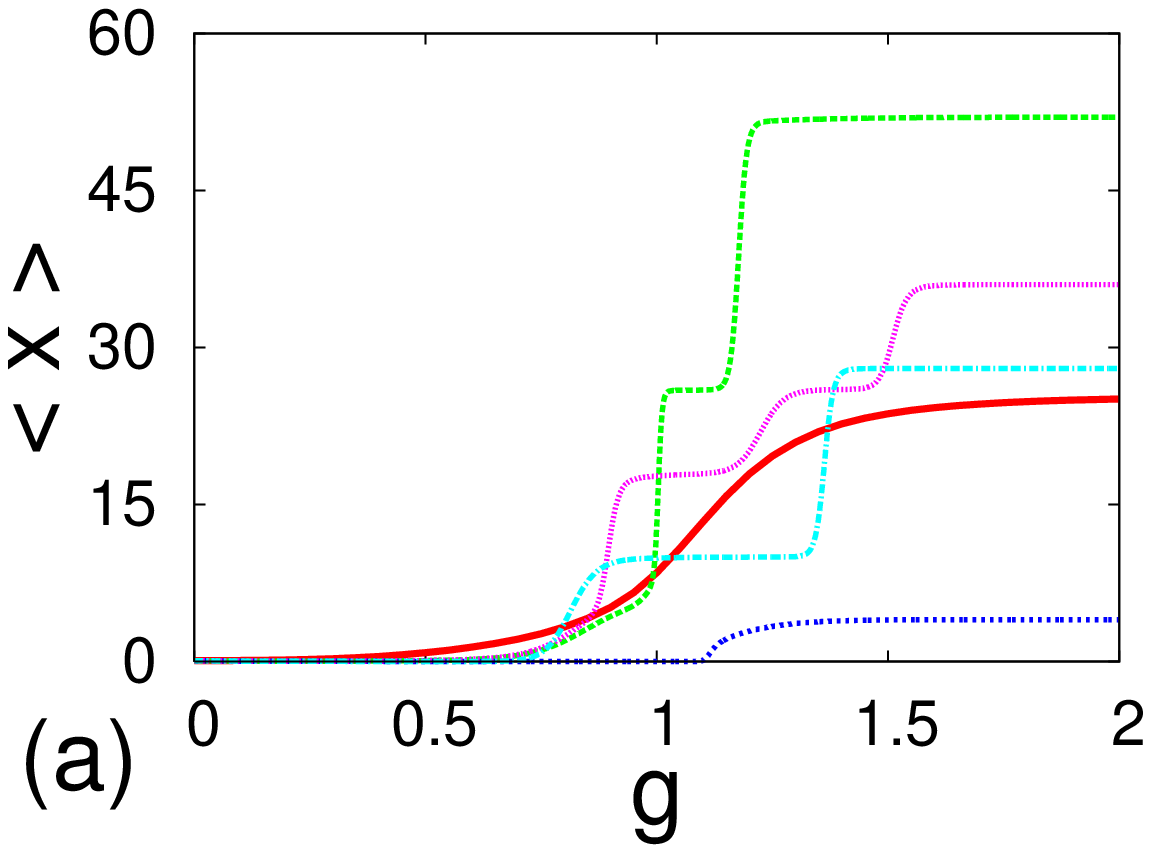}
		\includegraphics[width=1.6in]{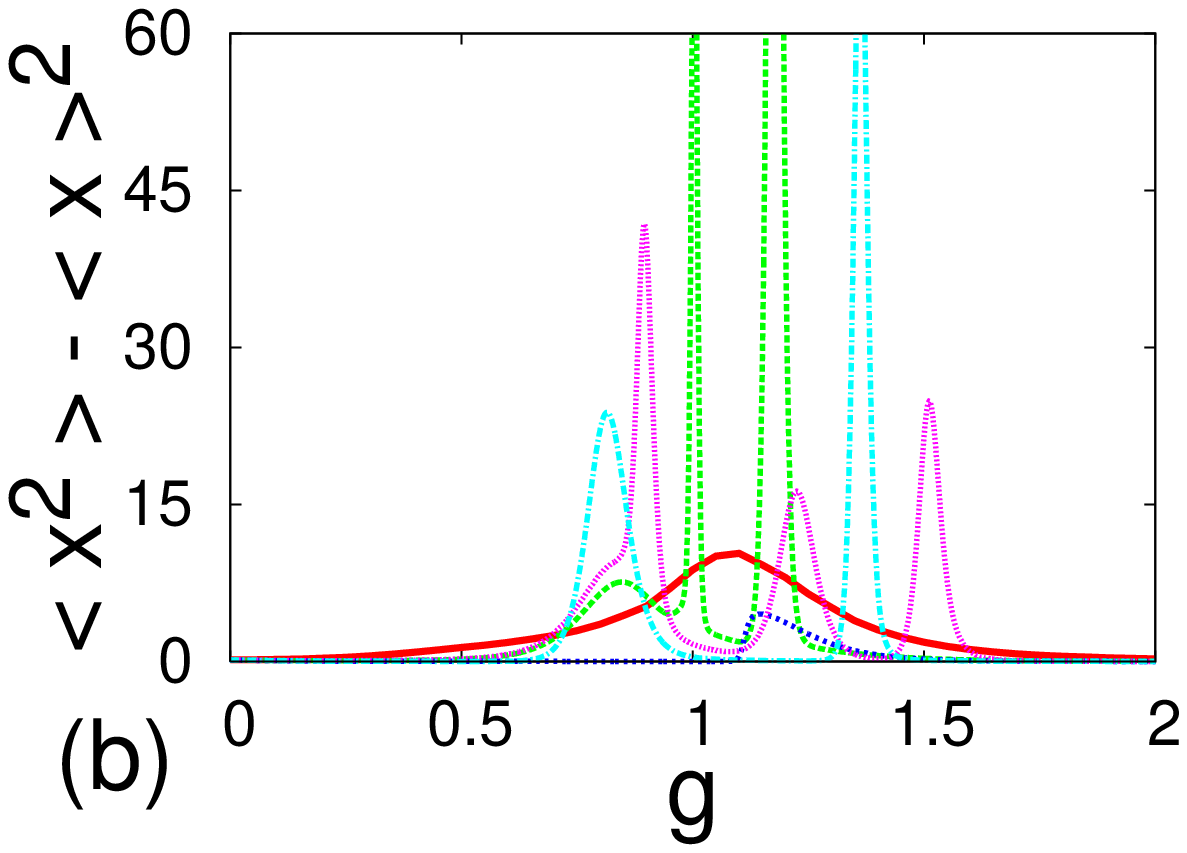}
        \includegraphics[width=1.65in]{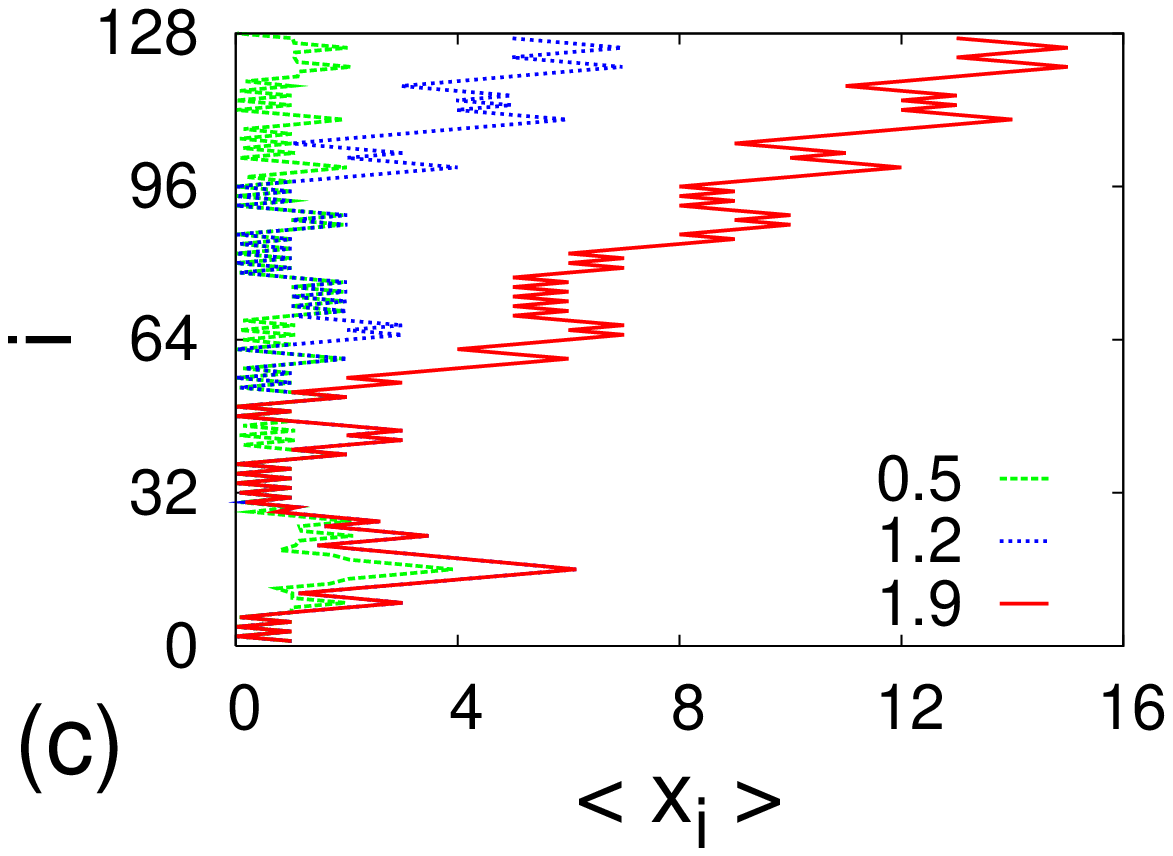}
		\includegraphics[width=1.65in]{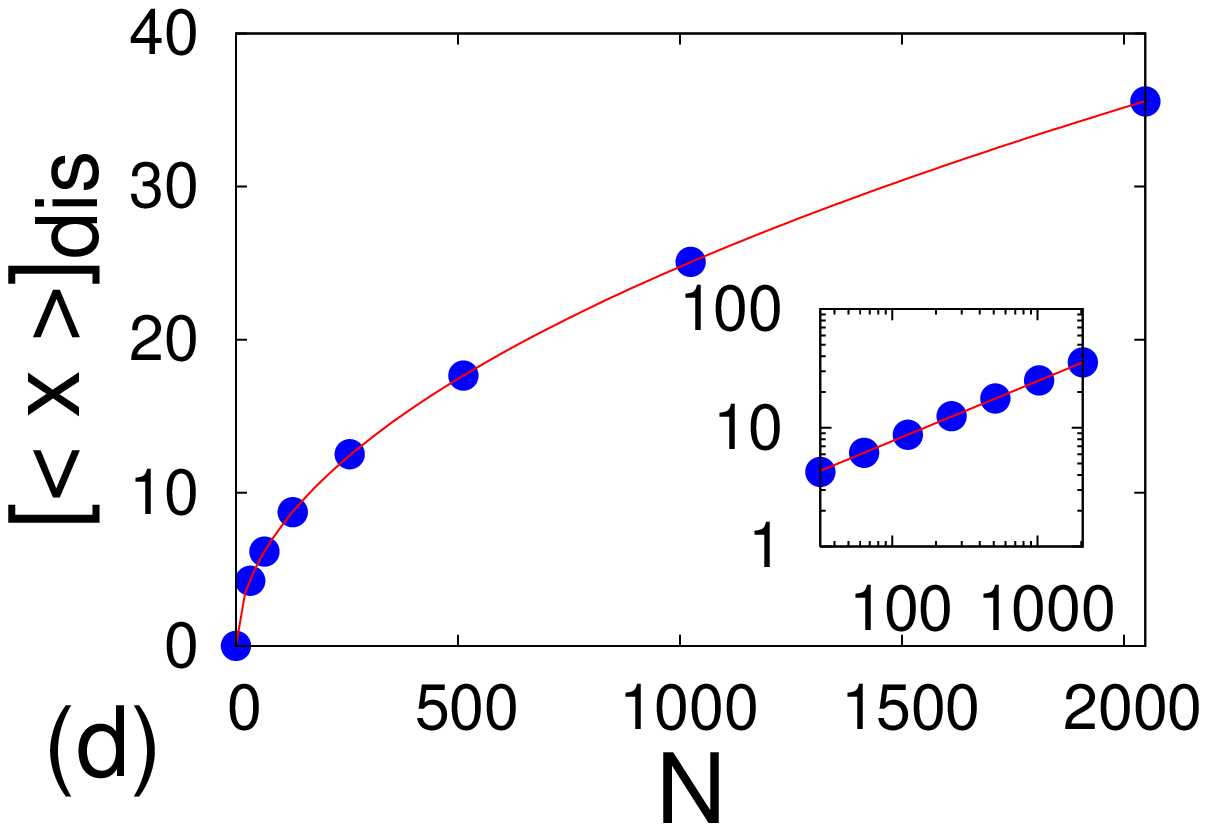}
		\caption{\label{fig:gx} (a) $\langle x \rangle$ vs g, (b)
		extensibility for four different samples of length $N=1024$ and
		the average over $10^5$ such samples (thick solid lines).(c)
		Typical configurations for $N=128$ for three different $g$
		as indicated. (d) $\left[ \langle x
		\rangle \right ]_{dis}$ vs $N$ for $g=2\epsilon$. The solid line
		is the best fit to the data. All at $T=0.3$ with $\epsilon=1$. }
	\end{figure}
}

\newcommand{\figext}
{\begin{figure}[htbp]
        \centering
        \includegraphics[width=1.65in]{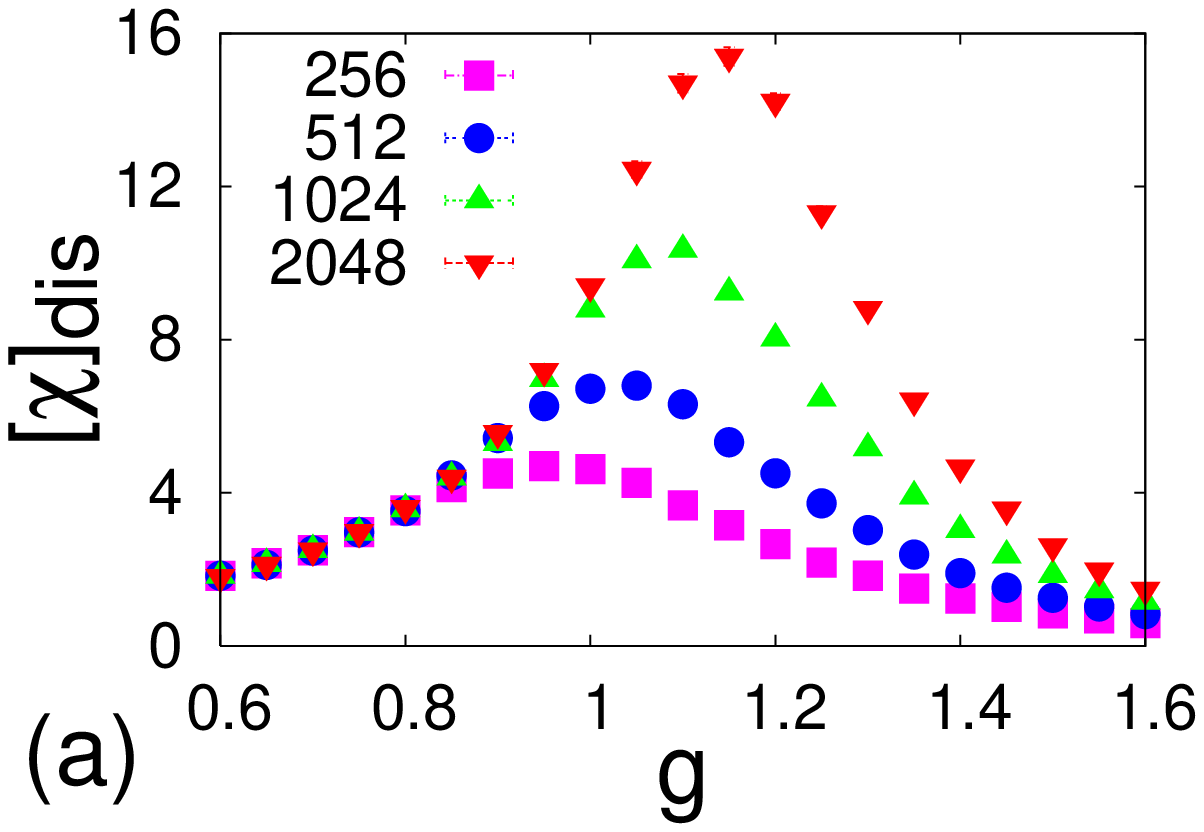}
        \includegraphics[width=1.65in]{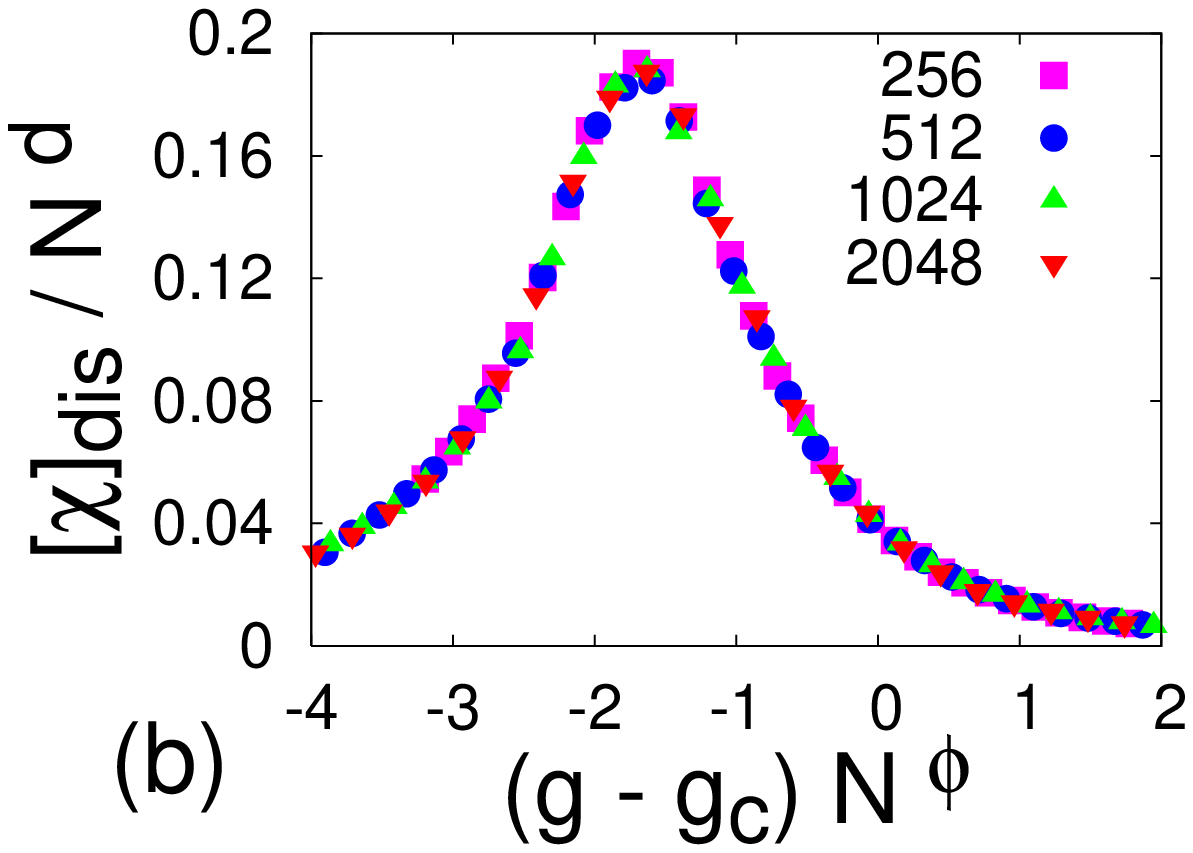}
		\caption{\label{fig:g0} (a) Extensibility $[ \chi ]_{dis}$ vs $g$ at
		$T=0.3$ with $\epsilon=1$ for various chain lengths. (b) Data
		collapse of the extensibility. 
		}
\end{figure}
}

\begin{document}

\title{Randomly forced DNA}
\author{Rajeev Kapri$^{1}$ and Somendra  M. Bhattacharjee$^{1,2}$}
\email{rajeev@iopb.res.in, somen@iopb.res.in}
\affiliation{ $^1$Institute of Physics, Bhubaneswar 751005 India.\\
	$^2$Saha Institute of Nuclear Physics, Kolkata 700064 India.}
\date{\today}
\begin{abstract}
        We study the effect of  random forces on a double stranded
        DNA in unzipping the two strands, analogous to the problem  of an 
        adsorbed polymer under a random force.  The ground state
         develops  bubbles of various lengths as the random force
        fluctuation is increased. The unzipping phase diagram is shown
        to be drastically different from the pure case.
\end{abstract}
\maketitle

Biological processes such as DNA replication and RNA transcription get
initiated and then proceed by unzipping of double stranded DNA (dsDNA)
by various enzymes like helicase, polymerase etc.\cite{watson}. These
enzymes exert force on dsDNA often directly, but also indirectly by
maintaining a fixed distance between the DNA strands (a fixed distance
ensemble). It was predicted theoretically that a dsDNA unzips to two
single strands (ss) when the force exceeds a critical value which
depends on
temperature\cite{somen:unzip,sebastian,maren:phase,scaling,kafri,tkachenko}.
The unzipping transition have also been studied experimentally by using
single molecule manipulations\cite{micro}.  A consequence of this force
induced unzipping transition is that in a fixed distance ensemble, a
dsDNA shows a coexistence of a zipped (ds) and an unzipped phase, known
as a Y-fork in biology, with a ``domain wall'' separating the two
phases as the junction of the Y. The motion of the domain wall under a
local instability either by a direct force or by the motor action of the
enzyme leads to a gradual nonequilibrium unzipping (as a propagating
front)\cite{helicase}. However for RNA polymerase or even dnaA helicase,
there is an additional need to open up a bubble at the right place
(``origin'' for initiation).  Though some force induced mechanism is
expected here, but not much details are known.  

The unzipping of DNA is a competition between the binding of the base
pairs (to be called monomers) and the orientation of individual links
connecting the monomers. A force applied at any point (say the end
point) gets transmitted to individual bonds to orient each one in the
direction of the force.  However, for a real DNA, there are various
sources of inhomogeneities.  Commonly studied cases of  sequence
heterogeneity, stacking energy, etc. generally affect the bound or the
ds part of the DNA.  In a cellular medium,  there are single strand
binding (SSB) proteins  which bind to single strands.  Such bound
objects  can lead to variation in the response of individual links (or
bonds) to the external force.  We study the role of such binding
proteins in DNA unzipping by considering a situation where the nature of
binding is modeled by a randomly oriented force. To avoid many
independent randomness, we avoid other heterogeneities like sequence
heterogeneity\cite{hetro}.  We compare our results with the DNA pulling
at the end.

Polymers with various types of randomness or disorder constitute a
special class of disordered problems because of the occurrence of
non-symmetry related (i.e., configurationally distinct) degenerate
ground states\cite{bkc05}.  The barriers (e.g., in space, in energy)
separating these states, the widths of the local wells, etc., then
determine the equilibrium and also the dynamic behavior of the polymer.
The random problem we discuss here is not one of those studied earlier,
but rather shows certain unique features, notably degeneracy of the
ground state at special points. All these features, biological
motivations apart, make this random force problem stand out on its own.
A close cousin is the problem of a directed polymer in a random medium.
But this problem is controlled by a ``$T = 0$'' fixed point (in a
renormalization group sense) in $D=1+1$ but a disorder induced phase
transition occurs in higher dimensions, $D>3$.  In the present case we
shall see a disorder-induced transition, even in $D=1+1$ dimension.

The strands of the DNA are complementary to each other and every base
in one strand knows its pair on the second strand. Considering the
case where the randomness works similarly on the two strands (i.e.,
either trying to keep the strands closer or unzip), the Hamiltonian in
the continuum can be written as\cite{somen:unzip}
\begin{eqnarray}
	\label{eq:1}
 	H_2 &=& \int_0^N \left [ \frac{1}{2}\left( \frac{\partial {\bf
	r}_1}{\partial z}\right )^{2} + \frac{1}{2}\left( \frac{\partial
	{\bf r}_2}{\partial z}\right )^{2} + V({\bf r}(z)) \right ]dz
	\nonumber\\
	& & \qquad \qquad \qquad -\int {\bf g}(z)\cdot \left( \frac{\partial
	{\bf r}_1}{\partial z} - \frac{\partial {\bf r}_2}{\partial
	z}\right) dz, 
\end{eqnarray}
where ${\bf r}_i(z)$ is the $d$-dimensional position vector of a monomer
at a length $z$ along the contour of the $i$th strand from one end
$z=0$, $N$ is the length of each strand or polymer, $V({\bf r})$ is the
binding potential, ${\bf r}(z)={\bf r}_1(z)-{\bf r}_2(z)$, and ${\bf
g}(z)$ is a random force.  Both the polymers are tied at end $z=0$. For
the ``pure'' problem, ${\bf g}(z)$ is constant and the force term reduces
to the standard form $-{\bf g} \cdot {\bf r}(N)$. The first two terms
on the right hand side represent the elastic energy or the connectivity
of each polymer (taken to be Gaussian). The base pair interaction is for
monomers at the same location on the two strands.  It follows that
an equivalent description can be obtained 
in which the two strands of the DNA are replaced by a relative
chain.  The Hamiltonian in terms of ${\bf r}(z)$is 
\begin{equation} \label{eq:2} 
	H = \int_0^N \left [ \frac{1}{2}\left( \frac{\partial {\bf
	r}}{\partial z}\right )^{^2} +  V({\bf r}(z))  -
	{\bf g}(z)\cdot \frac{\partial {\bf r}}{\partial z} \right ] dz,
\end{equation} 
with appropriate rescaling to make the elastic constant unity. 
This $H$ also describes
the problem of peeling of an adsorbed polymer by a pulling force
\cite{orlandini,kapri:pre05}.  Naturally, both the
problems have similar universal behaviour and many features of DNA
unzipping (at least qualitatively) can be understood by studying the
unzipping of an adsorbed polymer.  It is to be noted that if $z$ is
taken as an extra dimension (albeit different from the other $d$ spatial
coordinates), then these Hamiltonians also represent directed polymers
in $D=d+1$ dimensions.  This is the description we use in this paper.
We choose uncorrelated randomness with $[{\bf g} (z)]_{\rm dis} = 0 $
and $[g_i(z) g_j(z^{\prime})]_{\rm dis} =g^2 \delta_{ij} \delta(z -
z^{\prime})$.  Here, $[\cdots]_{\rm dis}$ denotes the quenched average
over force realizations.  We write ${\bf g}(z) = g \hat{\zeta}(z)$ where
the ${\hat{\zeta}}$ represents the random direction with unit variance.

The aim of the present paper is to study the effect of a random force of
zero mean on a bound or adsorbed polymer below its thermal unbinding or
desorption temperature.  If one is away from the DNA melting point or
the thermal desorption transition, then the characteristics of the
unzipping transition of the pure problem is not sensitive to the
dimensionality of the system, as seen explicitly in the exactly solvable
directed polymer problem in different dimensions including
$D=1+1$\cite{maren:phase,scaling}. With that in mind, we use the
discretized directed polymer model in $D=1+1$ dimensions.

\figmodel

The lattice version of Eq.~\ref{eq:2} is a polymer in $D=1+1$
dimensions, directed along the diagonal of a square lattice
(Fig.~\ref{fig:model}(a)). At $x=0$, there is an attractive,
impenetrable wall with binding energy $-\epsilon$ ($\epsilon>0$) which
favors adsorption of the polymer on the wall. For DNA this
impenetrability implies mutually avoiding chains. One end of the polymer
is always kept anchored on the wall while the other is left free. On
each bond between the two consecutive monomers, there is a random force
$g(z)=g \zeta(z)$ ($\zeta(z)$ same for a layer) which is always
perpendicular to the wall.  The magnitude $g$ related to the standard
deviation of the force is kept fixed but the direction $\zeta(z) = \pm
1$, i.e., either towards the wall or away from it, is chosen randomly
with equal probability so that the average force, $[g \zeta(z)]_{\rm
dis} =0$.  By averaging over the force configurations on the polymer
(quenched averaging), we find that even in the absence of a fixed
pulling force at the end, there is an unzipping transition if the
variance of the force fluctuation exceeds a critical value.  The
force-temperature phase diagram shows an increase of the critical force
with temperature (see Fig.~\ref{fig:model}(b)).  The critical force
starts decreasing only near $T_c$ and becomes zero at $T_c$.  This is to
be contrasted with the pure adsorption problem where the critical force
decreases monotonically with the temperature.

For every realization of the randomness, the partition function can be
calculated exactly as
\begin{equation}
	\label{pfn1d}
	Z_{n+1}^{\{\alpha\}}(x) = \sum_{j=\pm 1} Z_{n}^{\{\alpha\}}(x + j)
	e^{- j \beta g \zeta^{\{\alpha\}}(n)} {\mathcal W},
\end{equation}
where ${\mathcal W} = \left[1+(e^{\beta\epsilon}-1)\delta_{x,0}\right]$.
The superscript $\alpha$ in above expression denotes a particular
realization, and $Z_n(x)$ is the partition function for a polymer with
the $n$th (or the last) monomer at $x$.  Physical quantities are to be
averaged over the realizations. Quenched averaging is relevant 
here because the time scale of changes in the source of
randomness is much slower compared to the thermalization of the polymer.
One may note that an annealed averaging of Eq.~\ref{eq:1} would yield an
effective pure adsorption problem {\it without a force}\/ though with a
reduced elastic constant.  The quenched averaging is distinctly
different.

To monitor whether the polymer is zipped or unzipped, one needs the
average distance of the last monomer from the wall
($\langle\cdots\rangle$ denoting the thermal averaging) $[\langle x
\rangle]_{dis} =  [ { \sum_{x} x \,Z_N^{ \{ \alpha \}}(x)} / {\sum_{x}
Z_N^{ \{ \alpha \}}(x)} ]_{\rm dis}$ , and the isothermal extensibility,
being the response to the force, can be expressed in terms of position
fluctuation of the end monomer
\begin{equation}
	\label{eq:ext}
	\left [ \chi \right ]_{dis} \equiv \left [  \left.\frac{\partial
	\langle x \rangle}{\partial g}\right|_T\right]_{\rm dis} = (k_B
	T)^{-1} \left [ \langle x^2 \rangle - \langle x \rangle ^2
	\right]_{\rm dis}.
\end{equation}

In the presence of a fixed applied pulling force at the end, the
recursion relation can be solved
exactly\cite{maren:phase,scaling,orlandini,kapri:pre05}. In this case,
the critical force is  $g_c(T) = (k_B T/2) \ln \left [ e^{\beta
\epsilon} - 1\right ]$ with a temperature driven classical second order
desorption transition at $T_c = \epsilon / \ln 2$ for $g = 0$
\cite{kapri:pre05}(see Fig. \ref{fig:model}(b)). Henceforth $k_B$ is
absorbed in the definition of $T$ ($T\equiv k_B T$). The average
distance of the last monomer from the wall remains zero for any
$g<g_c(T)$ but becomes proportional to $N$ as soon as $g> g_c(T)$. This
shows that the force induced unzipping is a first order phase
transition.   There is no reentrance unlike the dsDNA problem because of
absence of zero temperature entropy of the bound state.

For the random problem, the ground state is the completely bound polymer
for zero force and new states can be obtained by flipping the adsorbed
monomers.  Fig.~\ref{fig:gs}(a) shows the four possible force
configurations of a monomer which is adsorbed on the wall. Let $n_1,n_2,
n_3$ and $n_4$ be the numbers of such configurations, then we have $n_1
+ n_2 + n_3 + n_4 = N/2$, since the geometry of the model permits only
$N/2$ monomers on the wall.   For small $g$, there is no gain in energy
in flipping. Only vertex for which we can gain is vertex {\it (iii)}, if
$ - \epsilon + 2 g > - 2 g$.  Therefore, if $g > \epsilon/4$, one
expects to see small bubbles. Below this force, the ground state is
unique. 
One may note that the last term of Eq.~\ref{eq:1} on integration by
parts contributes a fixed force at the end plus a force gradient which
acts locally on the chain. For a negative force gradient, the strands
minimize the free energy by maximizing the distance between them. Thus,
the bubble formation is not restricted to lattice models only and has
wider validity.  
For $g > \epsilon/4$, after flipping all the type {\it (iii)}\/
vertices, the average energy is
\begin{equation}
        \label{eq:gs}
        E_0 = -(n_1 + n_2 + n_4) \epsilon - 2 g (n_2 + n_3)
        = -\frac{N}{8} ( {3\epsilon} + {4g} ),
\end{equation}
taking all the four possible vertices to be equally probable.  Equating
this with the energy of the unzipped state, $-Ng$, in which the polymer
favors a configuration where each of the bond gets oriented in the local
force direction, we get the critical value of the force, $g_c = 3
\epsilon/4$. This analysis shows that there is a critical force
fluctuation above which the polymer favours the unzipped state at $T=0$.
Similar arguments would indicate that larger bubbles are possible for
$g>\epsilon/2$. These bubbles are different from the bubble (eye) phase
observed when the pulling force is applied in the interior of the
DNA\cite{kapri:dna}.

\figgs

By using the exact transfer matrix for the recursion relation of Eq.
\eqref{pfn1d} (and using log's to increase numerical accuracy), the
ground state energy can be determined from the free energy at very low
temperatures. By averaging over $10^5$ force realizations at $T=0.001$,
the estimated ground state energy per monomer for $N=256$ is shown in
Fig.~\ref{fig:gs}(b) for various values of $g$.  The same quantity is
also calculated using the single flip monte carlo at $T=0$ (squares) for
$N=1024$. Data from both the methods match excellently and are in
agreement with the prediction by the above analysis (Eq.~(\ref{eq:gs}))
for $ \epsilon/4 \le g \le 3\epsilon/4$.  The data also show that for $g
= \epsilon$, the free energy is $-Ng$, which is the energy of the
unzipped state. 

\figgx

A surprising feature of the model is that the ground state configuration
of the polymer depends on $g$ with degeneracies appearing when $g$ is an
integral multiple of $\epsilon/4$.  For example, for $g = \epsilon/4$,
flipping of vertex {\it (iii)}\/ does not cost any energy. On an
average, there are $N/8$ such two fold degenerate vertices.  Therefore,
the entropy of the ground state for $g=\epsilon/4$ is $S_0(g=\epsilon/4)
= (N/8) \ln 2$.  Similarly, if in a configuration, vertices {\it (i)}\/
and {\it (iii)}\/ are side by side, then flipping of these two vertices
do not cost any energy if $g = \epsilon /2$. Further opening of bubbles
would be possible by taking advantage of the forces on sites away from
the wall. One can similarly argue for other values of $g$. There is a
gradual increase of bubble sizes as $g$ is increased beyond
$g=\epsilon/4$. From the free energy we calculated the entropy at low
temperatures. After averaging over $10^6$ realizations and then
extrapolating to $T=0$ one gets the zero temperature or residual entropy
which for $g = \epsilon/4$ agrees nicely with the entropy calculated
above.

In order to explore the effect of thermal fluctuations, we study the
force-distance isotherms.  The $\langle x \rangle$ vs. $g$ isotherms for
four different samples of length $N=1024$ and also the average over
$10^{5}$ samples are shown in Fig.~\ref{fig:gx}(a).  The isotherm of an
individual sample shows steps similar (but different in origin) to the
steps seen in the force-distance isotherms when an adsorbed polymer is
subjected to a pulling force in a random environment\cite{kapri:pre05}.
The steps for the random medium case appear due to the pinning of the
polymer in the attractive pockets formed by the random distribution of
energy on the lattice sites with the force attempting to depin from
these pockets.  In the present case, there is a mutual competition
between the set of bonds with the force towards the wall and the others
with the force in opposite direction. The former set would like to opt
for a state with monomers on the wall while the rest trying to unzip
them.  The steps in the isotherm lead everytime to a comb-like
extensibility as shown in Fig.~\ref{fig:gx}(b).  This indicates that the
polymer responds in a ``jerky'' manner, by opening up local pockets of
pinned regions.  This is corroborated in Fig.~\ref{fig:gx}(c), where the
locations of individual monomers for a chain of length $N=128$ are
plotted for three different $g$ chosen from different plateaus of the
isotherm. It indeed shows adsorbed regions followed by unzipped regions
and  unzipping of adsorbed regions as $g$ increases.

%The average distance of the last monomer from the wall for the polymer
%of different lengths for $g= 2\epsilon$ at $T = 0.3$ (unzipped region)
%is plotted in Fig.~\ref{fig:gx}(d).The solid curve which is the best
%fit to the data shows that $\left [ \langle x \rangle \right ]_{dis} = a
%N^{\nu}$ with $a=0.753 \pm 0.006$ and $\nu = 0.505 \pm 0.001$.  This
%shows that in the unzipped state, the polymer stays at a distance of
%$\sqrt N$ from the wall and the configuration is {\it not} \/ like a
%directed polymer in a random medium for which $\nu=2/3$.  One may add
%that in the random medium problem the polymer gets fully stretched by
%the force in the unzipped state (i.e., depins from all the pockets).

Figure~\ref{fig:gx}(d) shows $\left [ \langle x \rangle \right ]_{dis}$
vs $N$ for $g= 2\epsilon$ at $T = 0.3$ (unzipped region). The solid
curve which is the best fit to the data shows that $\left [ \langle x
\rangle \right ]_{dis} = a N^{\nu}$ with $a=0.753 \pm 0.006$ and $\nu =
0.505 \pm 0.001$, i.e. the polymer stays at a distance of $\sqrt N$ from
the wall and the configuration is {\it not} \/ like a directed polymer
in a random medium for which $\nu=2/3$.  One may add that in the random
medium problem the polymer gets fully stretched by the force in the
unzipped state (i.e., depins from all the pockets).

The phase diagram is obtained from the disorder averaged extensibility.
Fig.~\ref{fig:g0}(a) shows the plot of extensibility $\left[ \chi
\right]_{dis} $ vs $g$, for the polymer of lengths $N=256,512,1024$ and
$2048$ at $T=0.3$ averaged over $10^5$ realizations. The growth in the
peak with $N$ indicates the possibility of a divergence in the
thermodynamic limit ($N \rightarrow \infty$). Its position can be
located using the finite size scaling of the form $[\chi]_{\rm dis} =
N^{d} {\mathcal Y}( (g - g_c) N^{\phi})$, where $g_c$ is the force at
which the discontinuity is located in the thermodynamic limit and $d$
and $\phi$ are the characteristic exponents.  The data collapse obtained
using the Bhattacharjee-Seno procedure \cite{bh:seno} is shown in
fig.~\ref{fig:g0}(b) which gives $d = 0.58 \pm 0.02$, $\phi = 0.22 \pm
0.01$ and $g_c = 1.46 \pm 0.04$, with $d/\phi \approx 2.6$, so that
$[\chi]_{\rm dis} \sim |g - g_c|^{-2.6}$. Similar collapse is also found
for $[\langle x \rangle]_{dis}$ (not shown). This indicates a continuous
transition though the exponent is different of that of the heterogeneous
DNA sequence.  We adopt this procedure at various temperatures to trace
out the $g-T$ phase diagram. For temperatures near $T_c$, lengths up
to $N=8192$ are used.

The phase diagram is shown in Fig.\ref{fig:model}(b). In contrast to the
pure case, there is an increase in $g_c(T)$ because of the loss of the
residual zero temperature entropy of the bound state\cite{scaling}.  

\figext

In conclusion, we have introduced and studied the problem of a randomly
forced DNA or its equivalent randomly forced adsorbed polymer. The
fluctuating force unzips the DNA by a gradual increase of bubble sizes.
For some discrete values of the force (integral multiple of $\epsilon/4$
in our case), the ground state of the random force problem is degenerate
and contains bubbles, which is in contrast to the pure problem where
there is only one ground state. It suggests the possibility of opening
up local bubbles in selective regions (proper pockets of force
distribution) without unzipping the whole DNA. Only experiments can tell
us if the initiation of DNA replication or RNA transcription is through
such a mechanism with the smaller molecules playing the role of random
force agents.

R.K. thanks SINP for hospitality.

\end{document}